\documentclass[twocolumn,showpacs,preprintnumbers,amsmath,amssymb]{revtex4}

\usepackage{graphicx}
\usepackage{dcolumn}
\usepackage{bm}
\begin{document} 

\title{Dynamical domain walls and spin-Peierls order in doped
       antiferromagnets: \\
       evidence from exact diagonalization of small clusters}
\author{R. Eder$^1$ and Y. Ohta$^2$}
\affiliation{$^1$Institut f\"ur Festk\"orperphysik, Forschungszentrum
  Karlsruhe, 76021 Karlsruhe, Germany\\
$^2$Department of Physics, Chiba University, Chiba 263-8522, Japan}
\date{\today}

\begin{abstract}
The hole-doped 2D t-J model is studied by exact diagonalization on a 
$5\times 4$ cluster which, unlike the standard tilted square
clusters, can in principle accomodate an antiphase domain wall. 
For  hole concentration $10\%$ and $J/t\ge 0.5$ the ground state
energy/site is lower than the conventional tilted square 
$\sqrt{20}\times\sqrt{20}$-cluster. 
In the ground state two holes form a loosely bound pair pinned to an
antiphase domain wall. The dynamical density correlation function shows 
sharp quasiparticle-like peaks, reminiscent of the `holons' in 1D 
chains, which suggest the existence of soliton-like propagating
domain walls. The dynamical correlation function of the bond-singlet
operator has a low-energy peak structure characteristic of
columnar spin-Peierls order, the dynamical spin correlation function
shows an intense and isolated `resonance peak' near $(\pi,\pi)$.
\end{abstract} 

\maketitle

The presence of stripe-like structures\cite{Tranquada}
which were predicted early on by mean-field studies
of Hubbard-like models\cite{hf},
is the basic idea underlying much of the recent theoretical
work\cite{Loew,TsunetsuguTroyerRice,Zaanen,VojtaSachdev,CastroNeto} 
on cuprate superconductors.
An interesting question is whether the `pure' 2D t-J model, which
still represents the simplest theoretical description of the
CuO$_2$ planes, develops any stripe-like structures, either in its
ground state or in low-energy excited states.
The most conclusive evidence for the spontaneous formation of such stripes
in the 2D t-J model is probably the density-matrix renormalization
group (DMRG) study of
White and Scalapino \cite{WhiteScalapino} and the related work
by Martins {\em et al.} \cite{Martinsetal}.  Exact diagonalization (ED)
of small clusters\cite{Dagotto} on the other hand, so far has not
produced really compelling evidence for hole stripes, at least in the
physical parameter regime. Evidence for stripe-like hole-density 
correlations from ED calculations has been reported by
Prelovshek and Xotos\cite{PrelovshekXotos}, but only for
relatively large values of $J/t\ge 1.5$. This is astonishing,
because all effective interactions in the t-J model 
are expected to be short ranged, whence any strong tendency to form 
stripe-like structures should make itself felt even in small
lattices - unless it is suppressed by finite-size effects.\\
As already stressed by White and Scalapino\cite{WhiteScalapino}
a potential source of such finite-size effects are the
boundary conditions imposed by the cluster geometry. Most
studies to date  have been performed on tilted-square
clusters\cite{Dagotto} with
periodic boundary conditions (PBC), which are chosen such as
to accomodate the N\'eel-type order parameter in the undoped 
system. On the other hand, such a geometry is precisely the wrong one
when an antiphase domain wall is present in the ground state. The mere 
geometry of the cluster then
can frustrate the domain wall, thus artificially enforcing a
homogeneous ground state. An ideal system to check this
would be the $5\times 4$-cluster with PBC, which on one hand
is appropriate to accomodate an antiphase domain wall parallel
to the $y$-direction and on the other hand can be compared directly
to the tilted square $20$-site cluster. As will be demonstrated,
this cluster shows very
clear and unambigous evidence for a stripe-like domain wall not
only in the ground state, but also in the form of excited states 
corresponding to a soliton-like propagating domain wall.
The  2D t-J model thus has a very strong intrinsic
tendency to form stripes - if it is allowed to do so.
Moreover, the formation of stripes seems intimately related to
spin-Peierls-like columnar singlet order
\cite{VojtaSachdev,Sushkov_doped}.\\
The Hamiltonian of the t-J model reads
\[
H = -t \sum_{\langle i,j\rangle,{\sigma}}
\left( \hat{c}_{i,\sigma}^\dagger \hat{c}_{j,\sigma} + H.c.\right)
+ J\sum_{\langle i,j\rangle} 
\left({\bf S}_i \cdot {\bf S}_{j} - \frac{n_i n_j}{4}\right)
\]
Here $\langle i,j\rangle$ denotes summation over
nearest neighbor pairs,
$\hat{c}_{i,\sigma} = c_{i,\sigma} (1-n_{i,\bar{\sigma}})$ and
${\bf S}_i$ and $n_i$ denotes the operators of electron spin and
density at site $i$, respectively.
\begin{figure}
\includegraphics{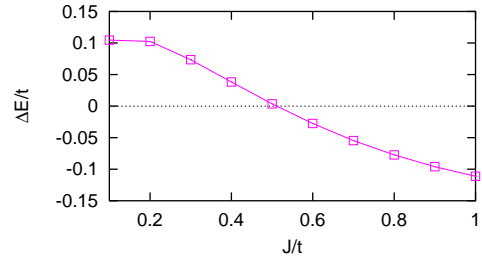}
\caption{\label{fig1} Difference $\Delta E_0$ between the
GS energy of the $5\times 4$ and 
the $\sqrt{20}\times\sqrt{20}$ cluster, as function of $J/t$.}
\end{figure}
We begin with a comparison of the ground state (GS) energy
for the $5 \times 4$ cluster and the tilted square $20$-site cluster.
One might expect that this can give an indication, as to
whether a given cluster has the proper geometry to describe the system.
Inappropriate boundary conditions introduce 
`frustration' thereby raising the energy. For example,
at half filling the GS energy in $5 \times 4$ is $-1.165 J/$site,
whereas in the square-shaped cluster it is $-1.191 J/$site,  the higher
energy for  $5 \times 4$ obviously being due to
the frustration of the (quasi-) N\'eel order along the odd-numbered
side of this rectangular
cluster. Interestingly enough, this changes when holes are added.
Figure \ref{fig1} compares the GS energy
of the two clusters with $2$ holes as function of $J/t$. 
For $J/t > 0.5$ the rectangular cluster indeed has the lower
energy, indicating that for the doped case the different boundary
conditions are at least as reasonable to represent the bulk system
as is the square cluster. All data to be presented below have been
obtained for the value $J/t=0.5$ and $2$ holes.
\begin{table}[b]
  \begin{center}
\begin{tabular}{|r r|rrr|rrr|}
\hline
&  &        & $g_D({\bf R})$ & &        & $g_S({\bf R})$ &  \\
\hline
&  2 & 0.298  & 0.139 &  0.035 &  0.127 & -0.110 &  0.032 \\
$R_y\uparrow$ &  1 & 0.239  & 0.144 &  0.041 & -0.299 &  0.110 & -0.040 \\
&  0 & 2.000  & 0.049 &  0.018 &  0.675 & -0.198 &  0.034 \\
\hline
&    &  0     &  1    &   2    &   0    &   1    &   2    \\
&    &        &  $R_x \rightarrow$ & &       & $R_x \rightarrow$ & \\
\hline    	     	           	          	   
\end{tabular}
\caption{Static correlation functions
for the $4\times 5$ cluster.}
\label{tab1}
\end{center}
\end{table}
Table \ref{tab1} shows the static density and 
spin correlation functions for the $5\times 4$ cluster with $2$ holes.
These are defined as the GS expectation values
$g_D({\bf R}) = \sum_j \langle n_{j} 
                 n_{j+{\bf R}} \rangle$ and
$g_S({\bf R}) = \frac{1}{N} \sum_j \langle {\bf S}_{j}\cdot
                  {\bf S}_{j+{\bf R}} \rangle$.
There is pronounced anisotropy in $g_D({\bf R})$ ,
particularly so at short distances, which immediately suggests a
`hole stripe' in $y$-direction.
The fact that the hole density correlation is stronly directional
might seem to suggest that this state is
related to the $p$-like pairing states, which usually form the first
exited state above the $d_{x^2-y^2}$-like ground state 
(and sometimes in fact the ground state itself)\cite{Symmetry}of two holes
in square clusters. These p-like pairs, however, are 
spin-triplets whereas the present ground state
is a spin singlet which moreover is even under reflection by both the
$x$- and $y$-axis. This state therefore is qualitatively very
different from those seen so far in square-shaped clusters.
Pronounced anisotropy is also seen
in the spin-correlation function, where in particular the
nearest neighbor spin correlation in $y$-direction (i.e. parallel
to the stripe) exceeds the one in $x$-direction by $50\%$.
While some anisotropy in the correlation functions is to be expected
solely due to the rectangular shape of the $5 \times 4$ cluster,
the edges differ in length by only
$20\%$ which seems to be rather small to explain the strong
anisotropy. \\
\begin{figure}
\includegraphics{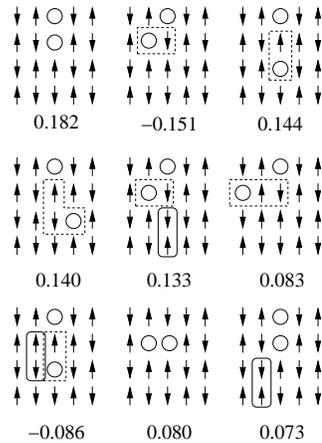}
\caption{\label{fig2} Hole-spin configurations of the representatives
with the largest weight in the GS wave function.
Dashed and solid boxes indicate the hops and spin flips
to created the state from the `seed state' in the top left panel.}
\end{figure}
To further clarify the nature of the ground state, Figure \ref{fig2}
shows the `representatives' of those basis states, which have the
largest weight in the GS wave function, labelled by their
coefficients in the GS. Taken together, these $9$ states
exhaust $14.1\%$ of the GS wave function - which is a lot,
given that there are all in all 59000 basis states for this
system. For clarity we mention that
a basis state for the ED scheme is created from the `representative' by 
a) translating it by all $N$ different lattice vectors of the cluster,
b) applying all $N'$ point group operations of the little group of the
total momentum ${\bf k}$ c) applying the
time reversal operator and d) adding up the resulting $2 N N'$ states
with the proper phase factors so as to create a state with prescribed
momentum, point group symmetry and parity under time reversal.
All states in Figure \ref{fig2} show a domain wall separating
two perfect N\'eel states. In the configuration with the largest
weight (top left) the two holes form a pair parallel to the domain wall.
All subsequent states can be generated from this 
`seed state' either by hole hopping, by a quantum spin flip along
the domain wall, or by a combination of both.
This is very much reminiscent of  `string picture' theories
for holes populating a domain wall between two
N\'eel states\cite{EderWrobel,Chernyshev} and in particular highlights
the importance of charge fluctuations transverse to the
stripe\cite{EderWrobel,Chernyshev}.
Taken together the data presented so far
show that the $5\times 4$ cluster has a qualitatively new
(as compared to the tilted square clusters)
ground state: a loosely bound hole pair pinned to an antiphase domain
wall of the staggered magnetization, with strong quantum spin fluctuations
along the fault line.\\
Having established the nature of the ground state, we consider the
excitation spectra of the system. As a first step, we address
the existence of dynamical domain walls.
Here the appropriate quantity to look at is the
dynamical density correlation function (DCF), defined as
\begin{equation}
D({\bf q},\omega) = \Im \frac{1}{\pi} \sum_\nu
\frac{ |\langle \Psi_\nu | n_{\bf q} | \Psi_0 \rangle
  |^2 }{\omega - (E_\nu-E_0) -i0^+}, 
\label{correlation_function}
\end{equation}
where $n_{\bf q} =  \frac{1}{\sqrt{N}}
\sum_{j} n_j e^{i{\bf q} \cdot {\bf R}_j}$ is the Fourier transform of the
electron-density operator $n_j$.
Figure \ref{fig3} compares $D({\bf q},\omega)$ for the
$5\times 4$ and the tilted square $20$-site cluster.
Whereas the DCF for the tilted square cluster
is essentially incoherent\cite{density_corr},
the DCF for the rectangular cluster shows intense
and well defined peaks at the lower edge of the spectra, particularly so
at  ${\bf q}=(2\pi/5,0)$ and ${\bf q}=(4\pi/5,0)$.
In fact, it is tempting to compare the DCF for the $4\times 5$ cluster 
to the one for 1D chains.  Figure \ref{fig3} also
\begin{figure}
\includegraphics{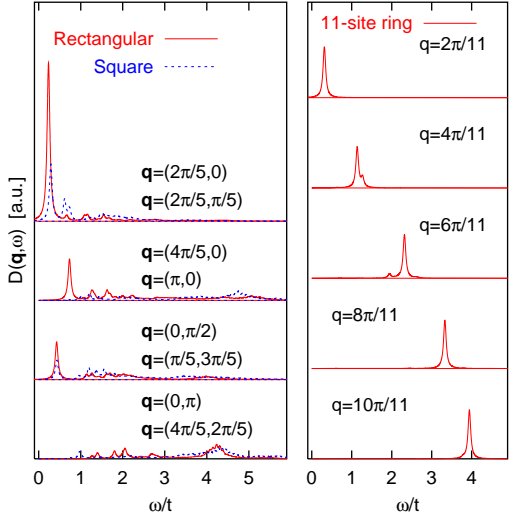}
\caption{\label{fig3} Left: DCF for the $4\times 5$ and 
$\sqrt{20}\times \sqrt{20}$
cluster. The upper/lower momentum label refer to  
$4\times 5$/$\sqrt{20}\times \sqrt{20}$.\\
Right: DCF for a single hole in a 1-D
$11$-site ring at $J/t=0.5$.}
\end{figure}
shows the DCF for
an 11-site t-J chain with one hole. The DCF here really is an almost
ideal single-peak spectrum, with a dispersion that
can be fitted  very accurately
by the expression $\epsilon(q)=2t(1+\cos(q))$. Obviously these peaks
originate from the particle-hole excitations
of the single `holon' in the system.
Since the holon is nothing but a domain wall, this suggests
that the sharp peaks in the $5\times 4$
DCF for momenta
$(q_x,0)$ also originate from the propagation of the domain wall as a whole.
The stronger ${\bf q}$-dependence of the {\em peak weight} in the 2D cluster
is probably due to the finite width of the
domain wall: whereas the holon really is a point-like 
object, making its density structure factor $g_D(q)$ flat in $q$-space,
the domain wall in 2D has some extension in $x$-direction, whence
its structure factor must have a $q_x$-dependence.\\
Next, we focus on the anisotropic spin correlations.
Read and Sachdev\cite{ReadSachdevI}  have
proposed a spontaneous spin-Peierls-like dimerization to occur
as a general feature of $S=\frac{1}{2}$ antiferromagnets in 2D - which
would explain the anisotropy of $g_S({\bf R})$. 
The appropriate quantity to check this hypothesis
is $B_\alpha({\bf q},\omega )$, the dynamical 
correlation function (defined as in (\ref{correlation_function})) 
of the bond-singlet operator $B_{\alpha,{\bf q}} = \frac{1}{\sqrt{N}}
\sum_j \left({\bf S}_j \cdot {\bf S}_{j+{\bf e}_\alpha}-\frac{n_j n_{j+{\bf e}_\alpha}}{4}\right)
e^{i{\bf q} \cdot {\bf R}_j}$. Thereby $\alpha\in[x,y]$ denotes the
direction of the singlet-bond.
Also of interest is
$S({\bf q},\omega )$, the dynamical correlation function
of the spin operator $S_{\bf q} =  \frac{1}{\sqrt{N}}
\sum_{j} S_j e^{i{\bf q} \cdot {\bf R}_j}$.
These spectra are shown in Figure \ref{fig4}.
To begin with, $B_\alpha({\bf q},\omega )$ in $5\times 4$
has a prominent low energy peak for both singlet directions,
$\alpha=x,y$, at ${\bf q}=(2\pi/5,0)$.
\begin{figure}
\includegraphics{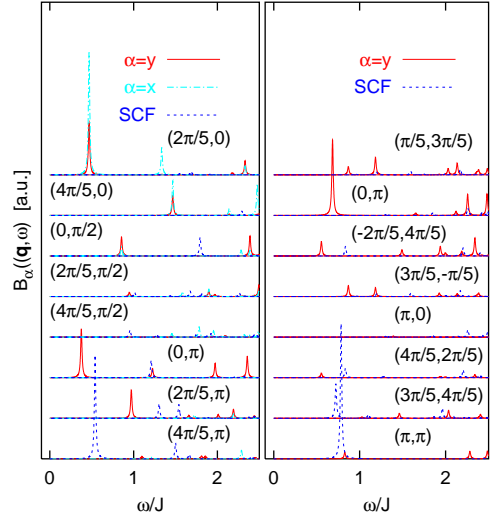}
\caption{\label{fig4} Left: $B_\alpha({\bf q},\omega )$
and $S({\bf q},\omega )$
in the $4\times 5$ cluster with two holes.
Right: The same correlation functions for the $\sqrt{20}\times \sqrt{20}$
cluster. $S({\bf q},\omega )$ is multiplied by 0.5. } 
\end{figure}
This peak (as well as the similar peaks at $(0,\frac{\pi}{2})$
and $(4\pi/5,0)$) is simply a replica of the intense low energy peak
seen at the same  ${\bf q}$ in the DCF. Just as the density operator
itself,  $B_{\alpha,{\bf q}}$ is a number-conserving spin singlet,
whence $B_\alpha({\bf q},\omega )$ probes exactly the same final state 
manifold as the DCF.
Since the bond singlet operator (partially) couples to
fluctuations of the electron density, $B_\alpha({\bf q},\omega)$
must pick up the signal from
the propagating domain wall. This interpretation is confirmed by the
fact, that the 
excitation energies $E_\nu-E_0$ of the respective peaks in the DCF and 
$B_\alpha({\bf q},\omega)$
agree to computer accuracy, i.e. these peaks originate from
the same final state $|\Psi_\nu\rangle$.
More interesting therefore is the second high-intensity and
low energy peak, at $(0,\pi)$. This peak appears only in
$B_y({\bf q},\omega )$, that means for bonds
parallel to the domain
wall. The wave vector $(0,\pi)$ for bonds in $y$-direction
corresponds exactly to
the Spin-Peierls order found by
Read and Sachdev\cite{ReadSachdevI} and proposed to be a general
feature of doped antiferromagnets
by various workers\cite{VojtaSachdev,Sushkov_doped}.
This peak has a lower excitation energy than the 
dominant peak in the spin correlation
function $S({\bf q},\omega )$ at $(\frac{4\pi}{5},\pi)$,
indicating that spin-Peierls
order is the most likely instability of the system.
$S({\bf q},\omega)$ itself
is quite remarkable, especially when we compare it to the
tilted square cluster. There $S({\bf q},\omega )$ shows 
a series of low energy peaks near ${\bf q}=(\pi,\pi)$, which suggest
a relatively smooth dispersion with a shallow minimum
at the incommensurate wave vector $(\frac{3\pi}{5},\frac{4\pi}{5})$.
 $S({\bf q},\omega )$ in $5\times 4$, on the other hand,
consists of a rather diffuse high-energy 
continuum and an isolated low-energy
peak at $(\frac{4\pi}{5},\pi)$. The dispersion of the spin
excitations (if it exists) must be rather discontinuous
near this momentum.
Such a spin-excitation spectrum is very much reminiscent
of the `resonance peaks' observed in neutron scattering
experiments on the superconducting state\cite{Mook}.
In the framework of the spin-Peierls scenario, the peak should be
interpreted as follows:
starting from a columnar spin-Peierls state with singlets in $y$-direction, 
a N\'eel-ordered state can be generated by condensation of 
(Bosonic) bond-triplet excitations\cite{SachdevBhatt}
with momentum ${\bf Q}=(\pi,0)$. 
In the present case of an antiphase domain wall, this should
be replaced by ${\bf Q}=(\frac{4\pi}{5},0)$.
The spin operator ${\bf S}_{(\frac{4\pi}{5},\pi)}$ then creates and
annihilates the condensed triplet-excitations\cite{Gopalanetal},
so that the intense low-enery peak at $(\frac{4\pi}{5},\pi)$ 
is simply the fingerprint of the (quasi)-condensate of triplets, which
creates antiphase N\'eel-like spin correlations out of the columnar
spin-Peierls state. To conclude our discussion of the spin dynamics,
we note that $B_{y}({\bf q},\omega)$ for the $\sqrt{20}\times \sqrt{20}$
cluster also does show a low-energy peak at $(0,\pi)$ - the tendency
towards spin-Peierls instability thus really seems to be quite 
general\cite{ReadSachdevI,VojtaSachdev,Sushkov_doped}.
Since the geometry of square cluster is unfavourable for
an antiphase domain wall, however, this does
not make itself felt as clearly as in the $5\times 4$ cluster.\\
In summary, it has been shown that a rectangular cluster of the t-J model 
whose geometry does not explicitely frustrate 
an antiphase domain wall,
shows rather different behavior than the tilted square clusters
used conventionally for exact diagonalization.
The GS in this cluster
shows the clear and unambiguous signatures of a stripe-like 
domain wall, a comparison of the GS
energies shows, that such rectangular clusters are
at least as well suited to describe the bulk system, as are the
square clusters.  Clearly, this does not prove the existence of
stripes in the thermodynamic limit - just as a square-shaped cluster
will tend to suppress a domain wall, the rectangular one will favour it -
but given the previous evidence for stripes found by White and 
Scalapino\cite{WhiteScalapino} on much larger systems, it is
quite plausible that the stripes in the rectangular cluster
have a similar structure.\\
Stripe-like structures in the rectangular cluster then
appear not only in the ground state -
rather, there is evidence for excited states where
apparently the finite total momentum of the state is carried by
a soliton-like propagating domain wall -  which is very much
reminiscent of the `meandering domain wall' scenario put forward by
Zaanen {\em et al.}\cite{Zaanen}. 
The data also show the presence of 
quasistatic columnar singlet order\cite{VojtaSachdev,Sushkov_doped}
in the system, which seems to be either a prerequisite for
or an immediate consequence of the stripe formation.
\begin{figure}
\includegraphics{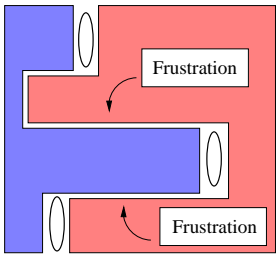}
\caption{\label{fig5} Mechanism for allignment of hole pairs.}
\end{figure}
Taken together, the above results suggest the
following scenario for doped antiferromagnets:
the spontaneous breaking of the point group 
symmetry of the lattice\cite{ReadSachdevI}
by formation of bond-singlet order produces ladder-like
patterns in the spin background of the system, which then
serve as `racetracks' for hole-pair-like solitons. Condensation of 
bond-triplets with a condensation amplitude that changes sign 
across a soliton
introduces strong antiferromagnetic correlations with opposite
staggered magnetization on the two sides
of the soliton. Under these circumstances, it becomes energetically
favorable for the solitons on different
ladders to form a line, because then the staggered magnetizations on the
different `ladders' can allign. Solitons which
propagate away from the stripe
create a track of magnetic frustration (see Figure \ref{fig5}), 
much as a single hole in an antiferromagnet, which creates an
effective potential that alligns the pairs.
In this way, one arrives at the
picture of a fluctuating domain wall\cite{Zaanen},
made of loosely bound hole pairs in a background provided
by the columnar singlets. An alternative point of view,
which also fits the data very well (see especially Figure \ref{fig2}), 
would be the accumulation of hole pairs
along the fault line separating two N\'eel states
with oppposite staggered magnetization, as
proposed early on by White and Scalapino\cite{White_Scalapino_unpublished}.

\end{document}